\documentclass[floatfix,aps,prd,twocolumn,showpacs]{revtex4} 

\usepackage{graphicx}

\usepackage{amsmath}
\usepackage{amssymb}

\font\FermiSmallfont=cmssq8 scaled 1200

\def\LANLppthead#1#2{
\null 
\begin{center}\vskip -1.0truein{\hbox to 7.5truein {
\hfill
\vbox to 1in {\vfill \FermiSmallfont
              \hbox{#1}
              \hbox{#2}
              \vfill}
}}\vskip-0.0truein\end{center}}

\def\rtilde{{\tilde{r} } }

\begin{document}

\LANLppthead {LA-UR 06-2765}{astro-ph/0605271}

\title{Constraints on Sterile Neutrino Dark Matter}

\author{Kevork Abazajian}
\affiliation{T-8 \& T-6, Theoretical Division, MS B285, Los Alamos National
  Laboratory, Los Alamos, NM 87545, USA }

\author{Savvas M. Koushiappas} 
\affiliation{T-6, Theoretical Division,
  \& ISR-1, ISR Division, MS B277, Los Alamos National Laboratory, Los
  Alamos, NM 87545, USA }


\begin{abstract}
  We present a comprehensive analysis of constraints on the sterile
  neutrino as a dark matter candidate.  The minimal production
  scenario with a standard thermal history and negligible cosmological
  lepton number is in conflict with conservative radiative decay
  constraints from the cosmic X-ray background in combination with
  stringent small-scale structure limits from the Lyman-alpha forest.
  We show that entropy release through massive particle decay after
  production does not alleviate these constraints. We further show
  that radiative decay constraints from local group dwarf galaxies are
  subject to large uncertainties in the dark matter density profile of
  these systems.  Within the strongest set of constraints, resonant
  production of cold sterile neutrino dark matter in non-zero lepton
  number cosmologies remains allowed.
\end{abstract}

\maketitle

\section{Introduction}

The nature of the dark matter remains a fundamental problem in
cosmology and particle physics.  Much can be gained from the inferred
non-gravitational properties of the dark matter such as its decay,
annihilation, interaction cross-section with baryonic matter, and
kinetic properties~\cite{Bertone:2004pz}.  
A class of candidate dark matter particles with no standard model 
interactions, but couplings to the standard model neutrinos via their mass 
generation mechanism, are the sterile neutrinos.
Sterile neutrinos can be produced in the early universe
via non-resonant matter-affected oscillations~\cite{Dodelson:1993je},
or through a resonant mechanism if there exists a non-negligible
lepton asymmetry~\cite{Shi:1998km}.

Weak interaction singlets such as sterile neutrinos arise naturally in
most extensions to the standard model of particle physics.  Grand
unified theories commonly contain singlets, which can act as sterile
neutrinos, and can play a role in their mass-generation
mechanism~\cite{Brahmachari:1998kt}.  Several models contain light
singlets as sterile neutrinos, including left-right symmetric (mirror)
models~\cite{Berezhiani:1995yi}, supersymmetric axinos as sterile
neutrinos~\cite{Chun:1999cq}, superstring
models~\cite{Langacker:1998ut}, models with large extra
dimensions~\cite{Arkani-Hamed:1998vp,Abazajian:2000hw}, and
phenomenological models such as the $\nu$MSM~\cite{Asaka:2005an}.
Letting the mass and mixing angle between the sterile and respective
active neutrino be free parameters, as well as the lepton number of
the universe be free, the sterile neutrino can behave as hot, warm or
cold dark matter, with masses in the range $\sim 0.1 - 100\rm\ keV$
~\cite{Abazajian:2001nj}.  The abundance of sterile neutrinos and its
relationship to the mixing parameters is affected by the quark-hadron
transition, however, this relationship is well known for the standard
prediction of a cross-over transition~\cite{Abazajian:2002yz}.

In the standard non-resonant production mechanism, sterile neutrinos
are produced via a collision-dominated oscillation conversion of
thermal active neutrinos.  Deviations from a thermal spectrum in the
sterile neutrinos are produced due to the change in the primordial
plasma's time-temperature relation during production, dilution due to
the disappearance of degrees of freedom, the modification of the
neutrino thermal potential from the presence of thermal leptons, and
the enhanced scattering rate on quarks above the quark-hadron
transition~\cite{AbazajianProduction05}.  As such, the resulting dark
matter momentum distribution is suppressed and distorted from a
thermal spectrum.
  
Sterile neutrinos exhibit a significant primordial velocity
distribution.  This has the effect of damping inhomogeneities on small
scales and thus sterile neutrinos behave as warm dark matter (WDM).
Models with a suppression of small scale power have drawn attention
due to their potential alleviation of several unresolved problems in
galaxy and small scale structure formation~\cite{Bode:2000gq}.  Of
particular interest recently are the possible indications of the presence of
cores in local group dwarf galaxies, inferred from the positions of
central stellar globular
clusters~\cite{Goerdt:2006rw,Sanchez-Salcedo:2006fa} and radial
stellar velocity dispersions~\cite{Wilkinson:2006qq}.  The primordial
velocity distribution produces a limit to the maximum phase-space
packing of the dark matter, which, if attained, can produce a cored
density profile for a dark matter halo.  Conversely, this places a
robust limit---the Tremaine-Gunn bound---on the mass and phase space
of the dark matter particle from observed dynamics in galaxy
centers~\cite{Tremaine:1979we}.

Lighter mass WDM particles more easily escape gravitational
potentials, and therefore suppress structure on larger scales, which
can be constrained by the observed clustering on small scales of the
Lyman-alpha (Ly$\alpha$) forest.  Possibly the most stringent limits
on the suppression of power on small scales are placed by inferring
the small scale linear matter power spectrum from observations of the
Ly$\alpha$
forest~\cite{Narayanan:2000tp,Viel:2005qj,Abazajian:2005xn,Seljak:2006qw}.
The same flavor-mixing mechanism leading to the production of the
sterile neutrino in the early universe leads to a radiative
decay~\cite{Pal:1981rm}.  The decay rate increases as the fifth power
of the mass eigenstate most closely associated with the sterile
neutrino, and increases as the square of the mixing angle, producing a
lighter mass neutrino and mono-energetic X-ray photon at half the dark
matter particle mass.  X-ray observations can either detect or
constrain the presence of a line flux from surface mass densities of
dark matter on the sky~\cite{Drees:2000wi,Dolgov:2000ew}.  So far,
X-ray observations have placed upper limits on the particle mass and
mixing angle relation of the dark matter sterile neutrino with
observations of the cosmic X-ray background, clusters of galaxies,
field galaxies, local dwarf galaxies and the Milky Way halo
~\cite{Abazajian:2001vt,Boyarsky:2005us,Boyarsky:2006zi,Boyarsky:2006fg,Riemer-Sorensen:2006fh}.

Sterile neutrinos lighter than those that may be the dark matter can
also play a cosmological role as {\it hot} dark matter.  One or more
such light sterile neutrinos may be required to produce the flavor
transformation seen in the Los Alamos Liquid Scintillator Neutrino
Detector (LSND) experiment~\cite{Athanassopoulos:1997pv,Sorel:2003hf}.
Such sterile neutrinos would be associated with mass eigenstates of
order $1\rm\ eV$, and therefore much lighter than a warm or cold dark
matter sterile neutrino.  A light sterile neutrino of the type
required by LSND would be thermalized in the early
universe~\cite{DiBari:2001ua,Abazajian:2002bj}, and is constrained by
limits on the presence of hot dark matter from measures of large scale
structure~\cite{Dodelson:2005tp,Seljak:2006bg}.  Such limits can be
avoided if the LSND-type sterile neutrino was not thermalized due to
the existence of a small lepton number, though they nonetheless may be
produced resonantly~\cite{Abazajian:2004aj}.

\begin{figure}[t]
\includegraphics[width=3.3truein]{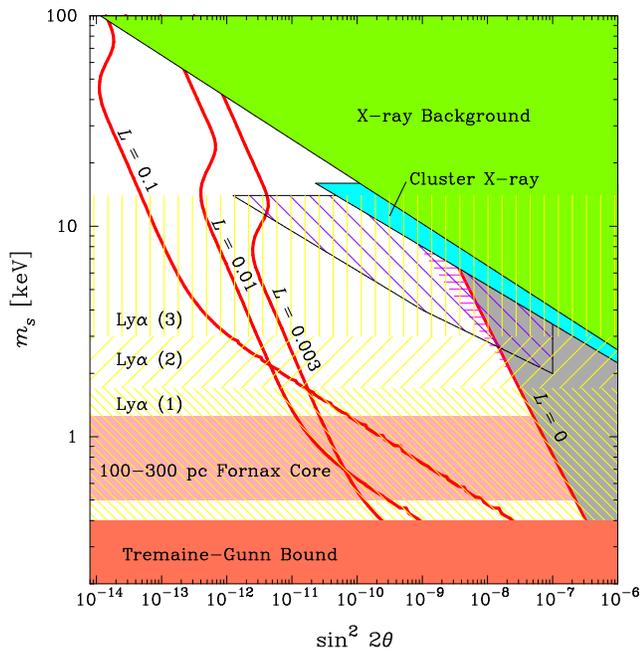}
\caption {\small Full parameter space constraints for the sterile
  neutrino production models, assuming sterile neutrinos constitute
  the dark matter.  Contours labeled with lepton number $L=0$,
  $L=0.003$, $L=0.01$, $L=0.1$ are production predictions for constant
  comoving density of $\Omega_s=0.24$ for $L=0$, and $\Omega_s=0.3$
  for non-zero $L$~\cite{Abazajian:2002yz}.  Constraints from X-ray
  observations include the diffuse X-ray background
  (green)~\cite{Boyarsky:2005us}, from XMM-Newton observations of the
  Coma and Virgo clusters (light blue)~\cite{Boyarsky:2006zi}.  The
  diagonal wide-hatched region is the claimed potential constraint
  from XMM-Newton observations of the LMC~\cite{Boyarsky:2006fg}.  The
  region at $m_s<0.4\rm\ keV$ is ruled out by a conservative
  application of the Tremaine-Gunn bound~\cite{Bode:2000gq}.  The
  regions labeled Ly$\alpha$ are those from the amplitude and slope of
  matter power spectrum inferred from the SDSS Ly$\alpha$ forest
  [Ly$\alpha$ (1)]~\cite{Abazajian:2005xn}, using high-resolution
  Ly$\alpha$ data [Ly$\alpha$
    (2)]~\cite{Viel:2005qj,Abazajian:2005xn}, and that from the
  high-$z$ SDSS Ly$\alpha$ of SMMT [Ly$\alpha$
    (3)]~\cite{Seljak:2006qw}.  The grey region to the right of the
  $L=0$ case is where sterile neutrino dark matter is overproduced.
  Also shown is the horizontal band of the mass scale consistent with
  producing a 100 - 300 pc core in the Fornax dwarf
  galaxy~\cite{Strigari:2006ue}. The parameters consistent with pulsar
  kick generation are in horizontal
  hatching~\cite{Kusenko:1998bk,Fuller:2003gy,Kusenko:2004mm}.
 \label{parameterspace}}
\end{figure}

There are two other interesting physical effects when sterile neutrinos 
have parameters such that they are created as the dark matter in the 
non-resonant production mechanism.
First, asymmetric sterile neutrino
emission from a supernova core can assist in producing the observed
large pulsar velocities above $1000\rm\ km\ s^{-1}$
\cite{Kusenko:1998bk,Fuller:2003gy,Kusenko:2004mm}.  The parameter
space overlaps that of the non-resonant production mechanism
(Fig.~\ref{parameterspace}). Second, the slow radiative decay of the
sterile neutrino dark matter in the standard production mechanism can
augment the ionization fraction of the primordial gas at high-redshift
(high-$z$)~\cite{Biermann:2006bu}.  This can lead to an enhancement of
molecular hydrogen formation and star formation, but also results in a
strong increase in the temperature of the primordial
gas~\cite{Mapelli:2006ej}.  This effect may have dire consequences on
the formation of the first stars, which remains an open
question~\cite{O'Shea:2006tp}.

In this paper, we review all of the constraints on the parameter space
of the sterile neutrino as a dark matter candidate, in conjunction
with the parameters needed for oscillation-based resonant and
non-resonant production mechanisms for the sterile neutrino as the
dark matter.  In \S\ref{constraints}, we review the best current
constraints on the sterile neutrino as a dark matter candidate.  We
review X-ray observation constraints in \S\ref{xray}.  The most
promising upper mass constraints come from X-ray observations of local
group dwarfs, but we show that they are subject to significant
uncertainties in the dark matter profile of the dwarf.  In
\S\ref{lya}, we discuss constraints from observations of the
Ly$\alpha$ forest. This provides 
the most stringent lower mass constraints, which
requires the free streaming length to be $\lesssim 90{\rm\ kpc}/h$, and
forces the primordial velocity dispersion to be too small to sustain
cored dark matter halos.  
With the combined constraints, we show that
the standard zero-lepton number non-resonant production model is
excluded if the most stringent constraints from the Ly$\alpha$ forest
are combined with the most conservative decay limits of the X-ray
background, and cannot be evaded in a model that dilutes and cools the
dark matter sterile neutrino by massive particle
decay~\cite{Asaka:2006ek}.  However, we show that the combined
constraints do not exclude resonant production mechanisms.

\section{Constraints}
\label{constraints}

The non-resonant ``zero'' lepton-number production calculation
presented in Ref.~\cite{AbazajianProduction05} is the simplest case
model for the production of sterile neutrinos as dark matter
candidates.  In this model, there are no extra couplings postulated
for the sterile neutrinos, and the cosmological lepton number is
comparable to the baryon number, and thus negligible.  The thermal
history (up to temperatures of $T\sim 500\rm\ MeV$) is given by
lattice QCD calculations through the quark-hadron transition, and
contains no sterile neutrinos in the initial conditions of the plasma,
due to the fact that the thermal potential suppresses sterile
production at high temperatures.

In Fig.~\ref{parameterspace}, we show contours of constant comoving
density comparable to the dark matter density for the non-resonant
zero lepton number ($L=0$) case, as well as enhanced resonant
production with initial cosmological lepton number cases ($L=0.003$,
$L=0.01$, $L=0.1$) from Ref.~\cite{Abazajian:2002yz}.  We have labeled
the standard prediction of $L\sim 10^{-10}$ as nil since it is
negligible for the non-resonant production mechanism. We define the
lepton number as
\begin{equation}
L \equiv \frac{n_{\nu} - n_{\bar\nu}}{n_\gamma},
\end{equation} 
where $n_\nu\ (n_{\bar\nu})$ is the number density of the neutrino
(antineutrino) flavor with which the sterile is mixed, and $n_\gamma$
is the cosmological photon number density.  The cosmological lepton
number is limited by the inferred primordial helium abundance and the
large to maximal mixing angle solutions to the solar and atmospheric
neutrino problems: $|L_e| < 0.05$, $|L_\mu+L_\tau| < 0.4$
\cite{Dolgov:2002ab,Abazajian:2002qx,Wong:2002fa}.  The constraints
discussed below are framed around the parameter space required for the
sterile neutrino dark matter production, and many are shown in
Fig.~\ref{parameterspace}.

\subsection{X-ray measurements}
\label{xray}

In this section, we review the sterile neutrino dark matter
constraints that come from measurements of the X-ray background, X-ray
measurements from the Virgo and Coma clusters, as well as measurements
of X-ray fluxes from the Draco local group dwarf.

It is straightforward to translate X-ray astronomy mass and mixing
angle parameter space constraints to sterile neutrino mass constraints
in the simplest model with the inversion of the production relation in
Ref.~\cite{AbazajianProduction05}:
\begin{equation}
\sin^2 2\theta = 7.31 \times 10^{-8} \left(\frac{m_s}{\rm
  keV}\right)^{-1.63} \left(\frac{\Omega_s}{0.26}\right)^{0.813},
\label{productionrelation}
\end{equation}
where $\theta$ is the mixing angle between the active and sterile
flavor states, and $\Omega_s$ is the fraction of the cosmological
critical density in sterile neutrinos.

If sterile neutrinos constitute the density associated with dark
matter, then their radiative decay would lead to a contribution to the
X-ray background (XRB hereafter)~\cite{Abazajian:2001vt}.  In a recent
analysis of the observed X-ray background from HEAO-1 and XMM-Newton,
Boyarsky et al.~\cite{Boyarsky:2005us} place the following limit on
the particle mass and mixing angle,
\begin{equation}
\sin^2 2\theta <  1.15 \times 10^{-4} \left(\frac{m_s}{\rm
    keV}\right)^{-5} \left(\frac{\Omega_s}{0.26}\right),
\label{xrbsin}
\end{equation}
with the corresponding exclusion region shown in
Fig.~\ref{parameterspace}.  If we combine this result with the
production Eq.~(\ref{productionrelation}), we find the corresponding
upper mass limit to be
\begin{equation}
m_s < 8.9\rm\ keV.
\label{xrbms}
\end{equation}

More stringent limits can be placed by X-ray observations of the large
dark matter surface mass density in clusters of
galaxies~\cite{Abazajian:2001vt}.  A recent analysis of XMM-Newton
observations of the Virgo and Coma clusters was presented in Boyarsky
et al.~\cite{Boyarsky:2006zi}. More specifically, it was shown that
near the center of the Virgo cluster (at radial distances $r < 11$
arcmin), the X-ray flux places a rough power-law constraint on the
$\sin^2 2\theta - m_s$ plane, as
\begin{equation}
\sin^2 2\theta < 10^{-2} \left(\frac{m_s}{\rm keV}\right)^{-6.64}.
\end{equation}
If this result is combined with the production mass-mixing angle
relation [Eq.~(\ref{productionrelation})], it results in a particle
mass constraint of \footnote{Abazajian, Fuller \&
  Tucker~\cite{Abazajian:2001vt} found a different limit due to a
  lower value of the central X-ray luminosity from the gas in
  Virgo. This had the effect of increasing the estimated signal to
  noise ratio. The resulting mass limit was therefore more stringent:
  $m_s < 8.2 \rm\ keV$, using the production relation
  Eq.~(\ref{productionrelation}).}
\begin{equation}
m_s < 10.6\rm\ keV.
\label{boyarskyvirgo}
\end{equation}

The combined Virgo and Coma analysis of Boyarsky et
al.~\cite{Boyarsky:2006zi} presents a more stringent limit. In this
case, an approximate power-law fit to their exclusion region places a
limit of
\begin{equation}
\sin^2 2\theta < 8\times 10^{-5} \left(\frac{m_s}{\rm
  keV}\right)^{-5.43},
\label{clustersin}
\end{equation}
which is considerably stronger than the XRB limit, Eq.~(\ref{xrbsin}).
Using the production relation Eq.~(\ref{productionrelation}), this
limit 
yields a sterile neutrino mass limit of 
\begin{equation}
m_s < 6.3\rm\ keV,
\label{clusterms}
\end{equation}
an improvement on the XRB limit, Eq.~(\ref{xrbms}).  

There are some notable issues with the Boyarsky et
al.~\cite{Boyarsky:2006zi} analysis of the Virgo and Coma cluster
data.  The analysis uses a fixed phenomenological model for the X-ray
emission of the cluster with specific lines added to fit atomic lines
in the spectrum.  On top of the phenomenological model, a Gaussian
line representing the potential sterile neutrino flux is inserted with
the width of the energy resolution of the instrument.  This method
does not allow modeling of the energy-dependent resolution of the
response of the detector, and can lead to a non-detection of a line
feature that exists at the position of an atomic line.  Emission lines
could be more properly modeled for the gas in clusters by using a
Mewe-Kaastra-Liedahl (MEKAL) model of the atomic and bremsstrahlung
emission of the gas~\cite{MEKAL}.  Nevertheless, barring the chance
coincidence of the sterile neutrino emission feature lying on an
instrumental feature or an atomic line, the limits,
Eq.~(\ref{clustersin}-\ref{clusterms}) from the Coma plus Virgo
analysis of Ref.~\cite{Boyarsky:2006zi} should be robust.

We now discuss the prospects of detecting sterile neutrino dark matter
in local dwarf galaxies.  It was recently proposed by Boyarsky et
al.~\cite{Boyarsky:2006fg} that X-ray observations by XMM-Newton of
nearby local group dwarf galaxies may present the best opportunity for
constraining or detecting the sterile neutrino decay flux of X-ray
photons.  The constraint region from that paper, using XMM-Newton
observations of the Large Magellanic Cloud (LMC), is shown as the
broad diagonal hatched region in Fig.~\ref{parameterspace}.

\begin{figure}[t]
\includegraphics[width=3.3truein]{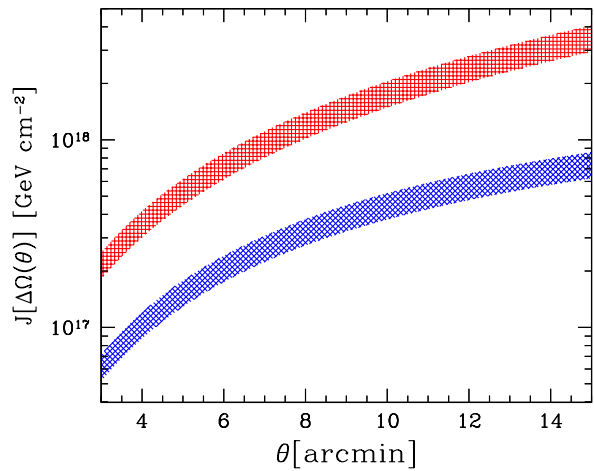}
\caption
{\small The value of the quantity $J [\Delta \Omega(\theta)]$ (see
  text) for two representative dark matter distributions in the Draco
  dwarf galaxy. The thickness of both curves corresponds to the range
  of values in each profile due to the distance uncertainties to
  Draco. The top curve corresponds to an NFW profile with
  $\log(\rho_s/M_\odot {\rm kpc}^{-3}) = 7.0$ and $\log (r_s/{\rm
    kpc}) = 0.85$, while the lower curve depicts a Burkert profile
  with $\log (\rho_s/M_\odot {\rm kpc}^{-3}) = 9.0$ and $\log( r_s
  /{\rm kpc})= -0.75$.
\label{draco}}
\end{figure}

Any analysis of the expected X-ray flux from sterile neutrino decays,
such as in Ref.~\cite{Boyarsky:2006fg}, is prone to the uncertainties
in the dynamical estimate of the dark matter distribution in the dwarf
galaxy~\cite{Booya}. For example, in the analysis of Ref.~\cite{MSC05}, the dark
matter profile of the Draco dwarf galaxy is consistent with both
cored and cusped dark matter distributions such as the NFW
~\cite{NFW96} and Burkert ~\cite{BUR95} profiles. These profiles may
arise in the case of massive ``cold'' sterile neutrinos (NFW) or for
lighter ``warm'' sterile neutrinos.  In order to demonstrate the
uncertainties in the X-ray flux due to the dark matter distribution in
Draco, we show in Fig.~\ref{draco} the value of the quantity $J[\Delta
  \Omega( \theta ) ]$ which is defined as the line of sight integral
of the matter distribution over a solid angle $\Delta \Omega(\theta)$
centered on the dwarf galaxy and expressed as
\begin{align}
\label{eq:Jdeltaomega}
J&\left[\Delta \Omega( \theta ) \right]\nonumber =\\\ \ \ &\quad
\rho_s \int_0^{2 \pi} d \phi \int_0^\theta \sin \theta '  
\left[ \int_{{x_{\rm min}(\theta ')}}^{{x_{\rm max}(\theta ')}} 
I[ \rtilde ( x  ) ] \, d x \right] d \theta ' .
\end{align}
Here, $I[ \rtilde ( x  ) ]$ is a function which depends on the assumed dark 
matter profile, and takes the form of 
\begin{eqnarray} 
I_{\rm NFW} \left[ \rtilde ( x  ) \right] &=& \frac{1}{ \rtilde( x )
  \left[ 1 + \rtilde( x ) \right]^2} \\   
I_{\rm BUR} \left[ \rtilde ( x  ) \right] &=& \frac{1}{ \left[ 1 +
    \rtilde(x) \right] \left[ 1 + \rtilde^2(x) \right] } ,
\end{eqnarray}
for the NFW and Burkert profiles respectively.  In
Eq.~(\ref{eq:Jdeltaomega}), $\rho_s$ is the characteristic density of
the assumed profile, the limits of the integration along the line of
sight are $x_{\rm max,min}\theta = D \cos \theta \pm \sqrt{ r_t^2 + (
  D \sin \theta )^2 }$, the quantity $\rtilde$ is defined as
$\rtilde(x) \equiv r(x) / r_s$, where $r_s$ is the scale radius of the
assumed profile, and it relates to the line of sight element through
$\rtilde(x) = \sqrt{x^2 + D^2 - 2 x D \cos \theta }$. The distance $D$
to Draco is $D=[75.8 \pm 7 \pm 5.4]\ {\rm kpc}$ estimated using RR
Lyrae variable stars ~\cite{BETAL03}, and the tidal radius $r_t$ is
taken to be $r_t \approx 7 \, {\rm kpc}$.  However, we point out that
as long as $r_t \gg r_s $ the dependence of $\Delta \Omega(\theta)$ on
$r_t$ is weak.

In Fig.~\ref{draco} we show the value of the quantity $J [\Delta
  \Omega(\theta)]$ for two representative cases of a cored and cusped
dark matter distributions. The top band depicts an NFW profile with
$\log(\rho_s/M_\odot {\rm kpc}^{-3}) = 7.0$ and $\log (r_s/{\rm kpc})
= 0.85$, while the bottom band corresponds to a Burkert profile with
$\log (\rho_s/M_\odot {\rm kpc}^{-3}) = 9.0$ and $\log( r_s /{\rm
  kpc})= -0.75$.  Both of these profiles are consistent with the
observed line-of-sight velocity dispersion measurements as shown in
~\cite{MSC05}.  Based only on profile and distance uncertainties, the
value of the quantity $\Delta \Omega(\theta)$ varies between $[2.5
  \times 10^{17} - 1.3 \times 10^{18}] {\rm GeV} \, {\rm cm}^{-2}$, a
factor of $5.2$ at 8 arcmin (which corresponds roughly to the XMM
field of view).

It has been claimed by Boyarsky et al.~\cite{Boyarsky:2006fg} that
perhaps the X-ray flux from the LMC could provide the strongest
constraint on the parameters of a sterile neutrino dark matter
candidate. Modeling the distribution of dark matter in the inner
regions of the LMC is even more uncertain due to the presence of a
stellar disk and a bar (as for the Milky Way as well).  As it was
shown in numerous studies, e.g., ~Ref.~\cite{vdMAHS02}, the LMC is
baryon dominated in the central region, with a mass-to-light ratio of
$\sim 3$ within the inner $\sim 9\rm\ kpc$ (for comparison, dwarf
spheroidals have mass-to-light ratios of $\sim 100$). The importance
of baryon domination on the distribution of dark matter in the LMC was
shown in the analysis of Ref.~\cite{AN00}. The derived mass of the LMC
is uncertain to within 20\%, depending on whether the disk is modeled
as ``maximal,'' or ``minimal.'' If the baryons are dominating, then
the rotation curve is much less sensitive to the distribution of dark
matter, making any estimate of the dark matter mass of the the LMC
unreliable.  In light of these uncertainties, we conclude that the LMC
is an unreliable Milky Way satellite for robust X-ray constraints from
detection of sterile neutrino decays.

Blank sky observations by X-ray telescopes may also provide a
detectable dark matter decay flux due to the dark matter halo of the
Milky Way itself~\cite{Riemer-Sorensen:2006fh,Boyarsky:2006fg}.  The
importance of uncertainties in the dark matter profile of the Milky
Way halo can also be significant.  Models of the measures of the
dynamics of the Milky Way can fit a range of halo masses of $(0.7 - 2)
\times 10^{12} M_\odot$~\cite{Klypin2001,Battaglia:2005rj}.  The
uncertainty in these dynamical estimates can lead to a factor of 3
difference in the expected X-ray flux in directions perpendicular to
the galactic plane, though these uncertainties are not reflected in
the analyses of Refs.~\cite{Riemer-Sorensen:2006fh,Boyarsky:2006fg}.

\subsection{Lyman-alpha Forest}
\label{lya}

The standard paradigm of cosmological structure formation is the
gravitational growth and eventual collapse from small to large scales
of initially adiabatic, Gaussian density fluctuations.  WDM particle
candidates alter the initial conditions of the perturbation spectrum
by damping small scale fluctuations below the free streaming scale of
the WDM.  The sterile neutrino particle mass is constrained from below
by the observations of small scale cosmological structure.

The most stringent lower bounds on the sterile neutrino mass arise
from observations of the clustering of gas along the line of sight to
distant quasars.  The density fluctuations of the gas follow that of
the dark matter to the scale at which the gas becomes pressure
supported.  The density fluctuations are linear to mildly nonlinear,
and can probe extremely small scale dark matter fluctuations.

Using a combination of cosmic microwave background observations, the
shape of the 3D power spectrum of galaxies from the Sloan Digital Sky
Survey (SDSS), the inferred linear matter power spectrum amplitude and
slope from the SDSS Ly$\alpha$ forest, the lower limit on the sterile
neutrino particle mass is $m_s > 1.7\rm\ keV\ (95\%\ CL)$.  In this
limit, a linear bias relation between the Ly$\alpha$ flux power
spectrum and matter power spectrum was
assumed~\cite{Abazajian:2005xn}.  Using the inferred matter power
spectrum from high-resolution spectra of the Ly$\alpha$ forest, the
limit was improved to $m_s > 3.0\rm\ keV\ (95\%\ CL)$; however,
significant systematic uncertainties exist in modeling the
high-resolution data~\cite{Abazajian:2005xn}. Seljak et
al.~\cite{Seljak:2006qw} (hereafter SMMT) find a much more stringent
constraint when directly using high-$z$ flux power spectra from the
SDSS measured by McDonald et al.~\cite{McDonald:2004eu} and other
higher-resolution flux power spectra:
\begin{equation}
m_s > 14\rm\ keV\ (95\%\ CL).
\label{seljaklower}
\end{equation}
All of these Ly$\alpha$ constraints are shown in
Fig.~\ref{parameterspace}.

Pioneering work on the sterile neutrino dark matter transfer function,
using the approximation of a suppressed thermal sterile neutrino
distribution, was done by Refs.~\cite{Colombi:1995ze,Hansen:2001zv}.
It has been shown that the sterile neutrino momentum distribution is
not simply a suppressed thermal distribution, but is significantly
nonthermal due to the effects on production of the dark matter by the
changing particle population in the early universe, lepton population
affecting the neutrino thermal potential, the quark-hadron transition,
and the dilution of the dark matter due to particle
annihilation~\cite{Abazajian:2001vt,AbazajianProduction05}.  All of
the above Ly$\alpha$ constraints use either the appropriate nonthermal
transfer function for sterile neutrinos modified by their production
at high temperatures in the early universe at or near the quark-hadron
transition~\cite{AbazajianProduction05}.  SMMT include an
approximation to these effects via an augmentation in the sterile
neutrino particle mass of 10\%, since the above effects cool the
momentum distribution.  The original assumption of a simple suppressed
thermal distribution for the sterile neutrino produces a suppression
scale that is altered by a factor of order 10\% in the particle mass
of the sterile neutrino, though the correction increases with sterile
neutrino particle mass due to their production at higher temperatures
where all of the above effects are more pronounced.  All of the
physical effects producing a nonthermal distribution where included in
the transfer function fit given in Ref.~\cite{AbazajianProduction05}.

The strength of the SMMT results arises from the sensitivity of the
high-$z$ Ly$\alpha$ flux power spectra to changes in the 1D linear
matter power spectrum, which is itself more sensitive to the
suppression scale of WDM than the 3D power spectrum.  At high-$z$, the
recovery of the amplitude of the power spectrum via nonlinear
clustering is reduced, enhancing the effects of WDM suppression.  The
temperature-density relation of the gas is constrained simultaneously
by the observations in SMMT, though the strong change in the thermal
state of the gas due to radiative sterile neutrino decay may be
significant~\cite{Mapelli:2006ej}.  Another essential feature of the
analysis in SMMT is the use of smaller volume hydrodynamical
simulations that can resolve the very small free-streaming scale of
$14\rm\ keV$ neutrinos.  The free streaming scale of a sterile
neutrino WDM is~\cite{Abazajian:2001nj}
\begin{equation}
\lambda_{\rm FS} \approx 840\ {\rm kpc}/h\left(\frac{\rm keV}{m_s}\right)
\left(\frac{\langle p/T\rangle}{3.15}\right),
\label{lambdafs}
\end{equation}
where $m_s$ is the mass state closely associated with the sterile
neutrino flavor, and $\langle p/T\rangle$ is the mean momentum over
temperature of the sterile neutrino distribution, and $h$ is the
Hubble parameter in units of $100\rm\ km\ s^{-1}\ Mpc^{-1}$.  A
thermal WDM particle is defined such that $\langle p/T\rangle/3.15
\approx 1$. Due to the thermal history of the universe during
production, the standard non-resonant production mechanism produces a
``cool'' sterile neutrino distribution, $\langle p/T\rangle/3.15
\approx 0.9$~\cite{AbazajianProduction05}, while the resonant
production mechanism enhances low-$p$ production, and $\langle
p/T\rangle/3.15 \approx 0.6$, depending sensitively on the mass of the
neutrino and initial lepton number~\cite{Abazajian:2001nj}.  The mass
scale within the free streaming length is
\begin{equation}
M_{\rm FS} \approx 2.6\times 10^{10}\  M_\odot/h\;\left(\frac{\Omega_{m}
h^2}{0.14}\right)\left(\frac{\rm keV}{m_\nu}\right)^3 \left(\frac{\langle p
/T\rangle}{3.15}\right)^3,
\label{mfs}
\end{equation}
where $\Omega_m$ is the fraction of the cosmological critical density
in matter.

To resolve the required $\lambda_{\rm FS}\approx 54{\rm\ kpc}/h$ of a
$14\rm\ keV$ neutrino, SMMT use a $20 {\rm\ Mpc} \,  h^{-1}$ box
with $256^3$ particles in dark matter and $512^3$ cells for gas,
providing a grid spacing of $39{\rm\ kpc}\, h^{-1}$.  This
allows only for a resolution of a fraction of the suppression due to
free streaming.  Higher resolution in principle should only enhance
the effects of the WDM suppression.  To test convergence, SMMT use a
single smaller volume simulation ($10{\rm\ Mpc}\, h^{-1}$) with
the same particle and grid spacing and find a $\sim$20\% change in the
magnitude of the effect, though no higher resolution simulations were
performed to test if the change is subsequently smaller as would be
expected in numerical convergence.  In addition, two other physical
effects are degenerate, to different extents, with the effects of WDM,
namely, pressure support due to the Jeans scale of the gas, and
temperature broadening.  These can mimic or hide the effects of the
free streaming scale of the WDM, therefore the thermal properties of
the gas are crucial to properly model in such limits.

A recent analysis by Viel et al.~\cite{Viel:2006kd} (herafter VLHMR)
used the same data of the SDSS Ly$\alpha$ flux power spectrum of
McDonald et al.~\cite{McDonald:2004eu}, but excluding higher
resolution data used by SMMT.  VLHMR utilizes a different method of
mapping the response of the flux power spectrum to changes in
cosmological and astrophysical parameters, {\it viz.}, VLHMR uses a
parameterized Taylor series expansion of the flux power response to
changes in physical paramters, with the Taylor parameters fit by
numerical simulations.  This method was shown to give similar results
for the standard $\Lambda$CDM cosmology~\cite{Viel:2005ha}.  VLHMR
finds a weaker sterile neutino particle mass lower limit, 10 keV (95\%
CL), than that of SMMT, 12 keV (95\% CL), using the Ly$\alpha$ data
from SDSS alone.  The discrepancy with SMMT is increased when taking
into account that the limit from VLHMR is too strong by $\gtrsim$10\%
because they do not use the correction for the ``cooler'' nonthermal
spectrum of the sterile neutrinos due to the effects in the early
universe described above.  Overall, between the two analyses there
exists a discrepancy of $\gtrsim 30\%$ in the limits on the sterile
neutrino mass.  This could arise due to different CMB and galaxy data
sets used: SMMT employs WMAP first year data and the SDSS galaxy power
spectrum; VLHMR employs WMAP third year data, higher resolution CMB
measurements and the 2dF galaxy power spectrum.  Therefore, it is not
certain whether the different CMB and galaxy data sets are the source
of the discrepancy, or whether it is in the hydrodynamical simulations
and method of mapping the response of the flux power spectrum to
changes in physical parameters.  However, both analyses find a
stringent limit due to the precision of the measurement of the
McDonald et al.~\cite{McDonald:2004eu} flux power spectrum at high
redshift.

To reflect back on one of the principle motivations for WDM, it is
important to note that it was shown by Strigari et
al.~\cite{Strigari:2006ue} that if the constraints from
Ref.~\cite{Abazajian:2005xn}, and especially SMMT are valid, then
dynamical constraints from the Fornax dwarf galaxies limit the size of
a core to be $\lesssim 85\rm\ pc$ and $\ll 10\rm\ pc$ for the two
Ly$\alpha$ limits, respectively.  In Fig.~\ref{parameterspace} we show
the particle mass required to produce a [100-300] pc core by inverting
the dynamical constraints of Strigari et al.~\cite{Strigari:2006ue}.
These constraints by Strigari et al. are based on the inferred dark
matter density profile from the radial velocity dispersion profile.
The positions of the globular clusters in Fornax may indicate a core
of $\sim$240 pc~\cite{Goerdt:2006rw,Sanchez-Salcedo:2006fa}.
Furthermore, the SMMT particle mass limit also limits the scale of the
suppression of the halo mass scale to be well below the typical masses
of dwarf galaxies, $M_{\rm FS} < 10^7 M_\odot\, {\rm h}^{-1}$.

In the resonant production model, the exact level of the lower bound
from the Ly$\alpha$ forest would be modified for each lepton number
case due to variation of $\langle p/T\rangle/3.15$ and therefore
$\lambda_{\rm FS}$ for each case.  However, this is at the level of
$\sim$30\% and is not monotonic across the lepton number region in
$m_s - \sin^2 2\theta$ space.  Therefore, we leave Ly$\alpha$ forest
as a horizontal line in Fig.~\ref{parameterspace}, to provide a rough
guide to the limit.  As shown in Fig.~\ref{parameterspace}, even with
the inclusion of all constraints, the resonant production model
remains unconstrained at high-mass scales: $14 \lesssim m_s \lesssim
100\rm\ keV$.

To summarize, the results of SMMT are extremely significant, ruling
out much of the parameter space that motivates WDM in general, and
when combined with conservative X-ray bounds, as shown above, they
rule out the standard sterile neutrino production mechanism.
Therefore, there is strong motivation to verify the robustness of the
SDSS Ly$\alpha$ measurements employed by SMMT as well as the modeling
by hydrodynamic simulations.

\section{Constraints in a Dilution Scenario}

It has been proposed that the production of sterile neutrino dark
matter could be followed by the decay of a massive particle, whose
decay products reheat the coupled species in the plasma, dilute the
sterile neutrino dark matter and cool it relative to the coupled
species~\cite{Asaka:2006ek}.  Though this involves a conspiracy
between the lifetime of the massive species and parameters coupling
the sterile neutrino to the active sector, it is an interesting
possibility that may alleviate structure formation constraints on the
sterile neutrino.  However, it is important to note that it does not
allow a window for {\it warm} dark matter, as this mechanism {\it
  cools} the WDM particle until it may be consistent with structure
formation limits.

In this scenario, a massive particle ($m\sim 100\rm\ GeV$) decays so
that the entropy release changes the relative abundance of the dark
matter by a factor $S$, i.e. $\Omega_s \rightarrow \Omega_s/S$.  In
the non-resonant oscillation production model, requiring a subsequent
increase in the mixing angle of the production relation to
\begin{equation}
\sin^2 2\theta = 7.31 \times 10^{-8}\ S^{0.813} \left(\frac{m_s}{\rm
  keV}\right)^{-1.63} \left(\frac{\Omega_s}{0.26}\right)^{0.813},
\label{productionrelation2}
\end{equation}
from Eq.~(\ref{productionrelation}). 
Using this and the conservative XRB limit, Eq.~(\ref{xrbsin}), the
limit on the entropy release factor is
\begin{equation}
S < 8.6\times 10^3 \left(\frac{m_s}{\rm keV}\right)^{-4.15}
\left(\frac{\Omega_s}{0.26}\right)^{-2.23}, 
\label{slimit}
\end{equation}
which is valid for the mass range $1 \le m_s \le 100\rm\ keV$.  With
extended models, $S$ can be as large as 100, at which point the
dilution is occurring for sterile neutrinos that were at or nearly at
thermal equilibrium with the plasma prior to the massive particle
decay.  Using Eq.~(\ref{slimit}) and the case where $S=100$, the limit
on sterile neutrino dark matter from the XRB is 
\begin{equation}
m_s^{\rm entropy} < 2.9\rm\ keV.
\label{xrbentropy}
\end{equation}

\begin{figure}[t]
\includegraphics[width=3.3truein]{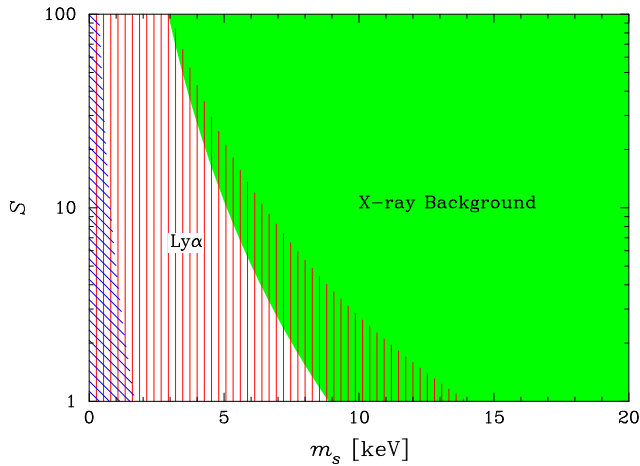}
\caption
{\small Shown here are the constraints on the massive particle decay
  dilution model.  The diagonally-hatched (blue) region is the
  lower-mass Ly$\alpha$ limit of Ref.~\cite{Abazajian:2005xn}, while
  the vertically (red) hatched region the Ly$\alpha$ limit of SMMT.
  In combination with the conservative XRB limit
  (green)~\cite{Boyarsky:2005us}, even extreme dilution models of
  $S=100$ are in conflict with combined constraints.  The standard
  case of no dilution corresponds to $S=1$.
\label{entropyrelease}}
\end{figure}

Limits from cosmological structure such as the Ly$\alpha$ forest are
also modified by the cooling of the sterile neutrino free streaming
length due to entropy release, such that the new lower mass limit is
\begin{equation}
m_s^{\rm entropy} = m_s^{\rm standard}\ S^{-1/3}.
\end{equation}
For the most stringent Ly$\alpha$ forest limit of SMMT,
Eq.~(\ref{seljaklower}), the entropy release model lower limit at
$S=100$ is
\begin{equation}
m_s^{\rm entropy} > 3.0 \rm\ keV.
\label{mlowerentropy}
\end{equation}
This is in conflict with the XRB in the entropy release model,
Eq.~(\ref{xrbentropy}).  The scaling of these relations is illustrated
in Fig.~\ref{entropyrelease}.  More stringent limits from X-ray
clusters or local group galaxies would be in stronger conflict with
Eq.~(\ref{mlowerentropy}).  Blank-sky observations including
contributions from the Milky Way halo also constrain high $S$
models~\cite{Riemer-Sorensen:2006fh}.  As such, we conclude
that massive particle decay does not open a new window for the
non-resonant oscillation production scenario.

\section{Conclusions}
\label{conclusions}

We have provided a comprehensive analysis of constraints on the
parameter space of interest for the non-resonant and resonant
production mechanisms of sterile neutrino dark matter.  Observations
in the X-ray of clusters of galaxies and the XRB place a limit to the
radiative decay rate of a sterile neutrino candidate and provide
constraints in the upper mass scale of the sterile neutrino dark
matter.  We have shown that limits from local group dwarf galaxies are
subject to large uncertainties in the dark matter profile of these
objects.  The lack of the effects of the suppression of small scale
power in the Ly$\alpha$ forest stringently limits the low mass scale
region of parameter space.

The SDSS Ly$\alpha$ constraints from the analysis of SMMT, when
combined with X-ray constraints, are in conflict with the sterile
neutrino being the dark matter in the standard non-resonant zero
lepton number production model.  The SMMT limits also exclude much of
the parameter space of interest for general WDM models.  Non-zero
lepton number cosmologies remain allowed for resonant production of
``cold'' sterile neutrino dark matter.  We find, however, that
dilution scenarios do not open a window for sterile neutrino dark
matter.  If the X-ray and Ly$\alpha$ constraints remain robust, then
only non-zero lepton number cosmologies remain viable for the
oscillation-production models of sterile neutrino dark matter.

\acknowledgments

We would like to to thank Peter Biermann, Alexey Boyarsky, James
Bullock, George Fuller, Katrin Heitmann, Lam Hui, Alex Kusenko, Julien
Lesgourgues, Maxim Markevitch, Uro\v s Seljak, Louie Strigari and John
Tomsick for useful discussions.  KA would like to thank the organizers
of the Sterile Neutrinos in Astrophysics and Cosmology 2006 Workshop,
where many of these discussions took place.  This work was supported
by Los Alamos National Laboratory under DOE contract W-7405-ENG-36.

\bibliography{sndm}

\end{document}